\documentclass[12pt]{iopart}

\usepackage{graphicx}
\usepackage{amssymb}
\usepackage{bm}
\renewcommand{\vec}[1]{\boldsymbol{#1}}
\newcommand{\beq}{\begin{equation}}
\newcommand{\eeq}{\end{equation}}
\newcommand{\beqa}{\begin{eqnarray}}
\newcommand{\eeqa}{\end{eqnarray}}
\newcommand{\w}{\omega}

\newcommand{\txt}{\mathrm}

\newcommand{\kv}{ {\bf k} }

\begin{document}

\title{Quantum dot Rabi rotations beyond the weak exciton-phonon coupling regime}
\author{Dara P. S. McCutcheon} 
\ead{dara.mccutcheon@ucl.ac.uk}
\address{Department of Physics and Astronomy, University College London, Gower Street, London WC1E 6BT, United Kingdom}
\address{London Centre for Nanotechnology, University College London, 17-19 Gordon Street, London WC1H 0AH, United Kingdom}
\author{Ahsan Nazir}
\ead{ahsan.nazir@ucl.ac.uk}
\address{Department of Physics and Astronomy, University College London, Gower Street, London WC1E 6BT, United Kingdom}

\date{\today}

\begin{abstract}

We study the excitonic dynamics of a driven quantum dot under the influence of a phonon environment, going beyond the weak exciton-phonon coupling approximation. By combining the polaron transform and time-local projection operator techniques we develop a master equation that can be valid over a much larger range of exciton-phonon coupling strengths and temperatures than the standard weak-coupling approach. For the experimentally relevant parameters considered here, we find that the weak-coupling and polaron theories give very similar predictions for low temperatures (below $30$~K), while at higher temperatures we begin to see discrepancies between the two. This is due to the fact that, unlike the polaron approach, the weak-coupling theory is incapable of capturing multiphonon effects, while it also does not properly account for phonon-induced renormalisation of the driving frequency. In particular, we find that the weak-coupling theory often {\it overestimates} the damping rate when compared to that predicted by the polaron theory. Finally, we extend our theory to include non-Markovian effects and find that, for the parameters considered here, they have little bearing on the excitonic Rabi rotations when plotted as a function of pulse area.

\end{abstract}


\maketitle

\section{Introduction}

Semiconductor quantum dots (QDs) provide a promising setting in which to explore the interplay of coherent control and decoherence in the solid-state. Spatial confinement of charge carriers gives rise to an atomic-like discrete energy level structure within the dot region~\cite{bayer00, banin99,bimberg98}, which allows for the selective probing of particular excitonic (electron-hole pair) transitions~\cite{bonadeo98,michler00}. This has lead to demonstrations of fundamentally quantum mechanical effects, such as laser-driven excitonic Rabi rotations~\cite{stievater01, kamada01, zrenner02, htoon02, borri02, ramsay10, ramsay10_2,muller07,flagg09,wang05,li03} and two-photon interference in QD emission~\cite{santori02,stevenson06,flagg10,patel10}. Moreover, optical preparation, control, and readout of a single self-assembled QD spin has been achieved~\cite{kroutvar04,mikkelsen07, berezovsky08, ramsay08,press08,atature06,gerardot08}, while various forms of coupling between closely spaced dots have been observed and characterised~\cite{bayer01,gerardot05, unold05,robledo08,stinaff06}.

Such experimental progress clearly demonstrates the feasibility of 
creating and manipulating both excitonic and spin quantum coherence in QD samples. However, despite this, QD charge carriers are often still strongly influenced by their surrounding solid-state environment. Though the resulting decoherence processes must generally be mitigated in order for QDs to be used, for example, in quantum information processing devices~\cite{loss98,imamoglu99, troiani00,biolatti00,piermarocchi02,pazy03,nazir04,economou08}, they also open up intriguing opportunities for exploring system-environment interactions in the solid-state. The combination of strong optical-dipole transitions, well developed control techniques, and relatively pronounced environmental interactions allows QDs to be used to study important open system effects that may be more difficult to observe, for instance, in atomic systems.

As an example, the damping of excitonic Rabi rotations in single self-assembled semiconductor 
QDs has been demonstrated to be {\it driving-dependent}~\cite{stievater01, kamada01, zrenner02, htoon02, borri02,ramsay10,ramsay10_2, wang05}. Owing to the large variation in QD type, growth strategy, and experimental set-up, a number of possible decoherence channels may be responsible for such behaviour. Two prominent mechanisms are off-resonant excitation of the wetting layer~\cite{wang05,vasanelli02,villas05}, and coupling to lattice vibrations (phonons)~\cite{ramsay10,ramsay10_2,machnikowski04, krugel05, vagov2007,nazir08}. In particular, recent experiments have provided compelling evidence that interactions with longitudinal acoustic (LA) phonons via deformation potential coupling dominates the damping of (ground-state) excitonic Rabi rotations in optically driven InGaAs/GaAs QDs~\cite{ramsay10, ramsay10_2}. Furthermore, while it might be hoped that certain decoherence sources could be suppressed by careful selection of samples and experimental techniques, ultimately self-assembled QDs are embedded in a host matrix. Interactions with phonons thus constitute an intrinsic limitation on the level of coherence seen in their excitonic transitions~\cite{machnikowski04,krugel05, vagov2007,nazir08,mahan, jacak03, krummheuer02, vagov02, grodecka07}. As such, a range of theoretical approaches have previously been developed to investigate the effects of phonon interactions on the coherent manipulation of excitons in QDs. Examples include perturbative expansions of the QD-phonon coupling, resulting in master equation descriptions of both Markovian~\cite{ramsay10, ramsay10_2, nazir08,erik} and non-Markovian~\cite{machnikowski04,alicki04,mogilevtsev08} nature, correlation expansions~\cite{krugel05,forstner03,krugel06}, and non-perturbative, numerically exact techniques which rely on calculation of the path integral~\cite{vagov2007}. 

The aim of the work presented here is to extend the master equation approach to QD exciton-phonon interactions beyond the weak-coupling regime in which it is commonly used~\cite{b+p}. The formalism is attractive because in many regimes analytical expressions are obtainable, and when they are not, it is not computationally expensive.
We present theoretical results describing the phonon-induced damping of a resonantly driven QD using a polaron transform~\cite{mahan,wurger98,wilson02} plus time-local master equation technique~\cite{b+p,jang08}. Our theory exploits a perturbative expansion in the polaron transformed representation, rather than in the system-bath interaction itself. As we shall show, under certain conditions this allows us to identify a perturbation term that is small over a much larger range of parameters than in the weak-coupling approach~\cite{wurger98}. In particular, our master equation is able to account for ``nonperturbative" effects not captured in a weak-coupling treatment, such as multiphonon processes and phonon-induced renormalisation of the driving pulse. This is particularly important in exploring the exciton dynamics at elevated temperatures (above $30$~K for the parameters we consider), where such effects may become important. Furthermore, we also extend the master equation to the non-Markovian regime. 

We focus particularly on comparing this theory to the weak-coupling Markovian technique used  in Refs.~\cite{ramsay10, ramsay10_2} to provide good fits to both the phonon-induced damping and energy shifts of the observed Rabi rotations. We find that for low temperatures (below $30$~K) the weak-coupling and polaron theories 
predict essentially the same excitonic dynamics, indicating that multphonon and renormalisation processes are unimportant. However, as the temperature is increased, we find surprisingly that the weak-coupling theory can {\emph{overestimate}} the phonon-induced damping rate when compared to the polaron approach. This is consistent with the weak-coupling fits to the highest temperature plots in Ref.~\cite{ramsay10_2}. We also show that, for the laser pulse durations used in Refs.~\cite{ramsay10} and~\cite{ramsay10_2}, the inclusion of non-Markovian effects within our formalism has an almost negligible influence on the Rabi rotations when plotted as a function of pulse area.

The paper is organised as follows. In Section~\ref{model} we introduce a model describing the QD system and its interactions with the phonon environment, and also define the polaron transformation. Section~\ref{master_equation} outlines our master equation derivation together with a discussion of the regimes in which we expect it to be valid. In Sections~\ref{time-dependent_driving} and~\ref{constant_driving} we consider resonant Rabi rotations, driven both by a Guassian laser pulse and constant driving, and compare the weak-coupling and polaron theory predictions. In Section~\ref{nonmarkov} we investigate non-Markovian effects, while in Section~\ref{summary} we give a brief discussion and summarise our results.

\section{Model and polaron transformation}
\label{model}

We consider a single QD modelled (as in Refs.~\cite{ramsay10,ramsay10_2}) as a two-level system with ground-state $|0\rangle$ and single-exciton state $|X\rangle$, separated by an energy $\omega_0$. The dot is driven 
by a laser of frequency $\omega_l$, with Rabi frequency $\Omega(t)$, and is coupled to a phonon bath 
represented by an infinite collection of harmonic oscillators with frequencies $\omega_{\bf k}$ and creation 
(annihilation) operators $b_{\bf k}^{\dagger}$ ($b_{\bf k}$). The system-plus-bath Hamiltonian takes the form (for $\hbar=1$)
\begin{eqnarray}
\label{Hamiltonian}
H&{}={}&\omega_0|X\rangle\langle X|+\Omega(t)\cos{\omega_l t}(|0\rangle\langle X|+|X\rangle\langle 0|)\nonumber\\
&&\:{+}\sum_{\bf k}\omega_{\bf k}b_{\bf k}^{\dagger}b_{\bf k}+|X\rangle\langle X|\sum_{\bf k}(g_{\bf k}b_{\bf k}^{\dagger}
+{g^*_{\bf k}}b_{\bf k}),
\end{eqnarray}
where the exciton-phonon couplings are denoted by $g_{\bf k}$. Moving to a frame rotating at the laser frequency $\omega_l$, and performing a rotating-wave approximation on the driving term, we obtain
\begin{eqnarray}
\label{HRWA}
H_{\rm RWA}&{}={}&\delta|X\rangle\langle X|+\frac{\Omega(t)}{2}(|0\rangle\langle X|+|X\rangle\langle 0|)\nonumber\\
&&\:{+}\sum_{\bf k}\omega_{\bf k}b_{\bf k}^{\dagger}b_{\bf k}
+|X\rangle\langle X|\sum_{\bf k}(g_{\bf k}b_{\bf k}^{\dagger}+{g^*_{\bf k}}b_{\bf k}),\nonumber\\
\end{eqnarray}
where $\delta=\omega_0-\omega_l$ is the detuning of the laser from the excitonic transition energy. The rotating-wave approximation can be justified here as both the Rabi frequency $\Omega(t)$ and detuning $\delta$ are generally small in comparison to $\omega_l$.

To move into the appropriate basis for the subsequent perturbation theory, 
we now apply a unitary polaron transformation to $H_{\rm RWA}$~\cite{mahan,wurger98}. This transformation displaces the bath oscillators when the QD is in its excited state; we shall explore in the following section how this can lead in many situations to a smaller perturbation term than a weak-coupling treatment of the exciton-phonon interaction in Eq.~(\ref{HRWA}). The transformed 
Hamiltonian is defined by $H_{\rm P}=e^{S}H_{\rm RWA}e^{-S}$, where 
\begin{equation}
\label{poltrans}
S=|X\rangle\langle X|\sum_{\bf k}(\alpha_{\bf k}b_{\bf k}^{\dagger}-\alpha_{\bf k}^*b_{\bf k}),
\end{equation}
with $\alpha_{\bf k}=g_{\bf k}/\omega_{\bf k}$. Hence, we may write 
\begin{equation}
e^{\pm S}=|0\rangle\langle0|+|X\rangle\langle X|\prod_{\bf k}D(\pm\alpha_{\bf k}),
\end{equation} 
where $\prod_{\bf k}D(\pm\alpha_{\bf k})=e^{\pm\sum_{\bf k}(\alpha_{\bf k}b_{\bf k}^{\dagger}-\alpha_{\bf k}^*b_{\bf k})}$ 
is a product of displacement operators $D(\pm\alpha_{\bf k})$. Defining the Pauli matrices in the $\{|0\rangle,|X\rangle\}$ basis as 
$\sigma_x=|X\rangle\langle 0|+|0\rangle\langle X|$, $\sigma_y=i(|0\rangle\langle X|-|X\rangle\langle 0|)$, 
and $\sigma_z=|X\rangle\langle X|-|0\rangle\langle 0|$, we find that our polaron-transformed Hamiltonian reads~\cite{wurger98,wilson02}
\begin{eqnarray}
\label{eqn:hpolpauli}
H_{\rm P}=\frac{\delta'}{2}\sigma_z+\frac{\Omega_r(t)}{2}\sigma_x+\sum_{\bf k}\omega_{\bf k}b_{\bf k}^{\dagger}b_{\bf k}+\frac{\Omega(t)}{2}\left(\sigma_xB_x+\sigma_yB_y\right),
\end{eqnarray}
where the detuning is now $\delta'=\omega_0'-\omega_l$, defined in terms of the 
{\it bath-shifted} QD transition energy $\omega_0'=\omega_0-\sum_{\bf k}\omega_{\bf k}|\alpha_{\bf k}|^2$, and we have 
ignored an irrelevant term proportional to the identity. Bath-induced fluctuations are now described by the Hermitian combinations
\begin{equation}\label{Bx}
B_x=\frac{1}{2}(B_++B_--2B),
\end{equation}
and
\begin{equation}\label{By}
B_y=\frac{1}{2i}(B_--B_+),
\end{equation}
where $B_{\pm}=\prod_\kv D(\pm \alpha_\kv)$, and $B=\langle B_{\pm} \rangle$ is the expectation 
value of the bath displacement operators. 

Importantly, the driving term in Eq.~({\ref{eqn:hpolpauli}}) has now 
been renormalised by a factor equal to this expectation value: $\Omega_r(t)=\Omega(t) B$. For 
a phonon bath in thermal equilibrium at inverse temperature $\beta=1/k_B T$, we find
\begin{equation}\label{Bathaverage}
B\equiv\langle B_{\pm}\rangle=\exp\Big[-(1/2)\sum_{\bf k}|\alpha_{\bf k}|^2\coth{(\beta\omega_{\bf k}/2)}\Big].
\end{equation}
For models of the type studied here, the system-bath interaction is entirely characterised by the spectral density $J(\w)=\sum_\kv |g_\kv |^2\delta(\w-\w_\kv)$~\cite{leggett87}. We are specifically interested in the coupling of bulk LA-phonons 
to our QD exciton, shown to dominate the dephasing dynamics in Ref.~\cite{ramsay10}, and therefore take a spectral density in the continuum limit of the form~\cite{ramsay10, ramsay10_2, krummheuer02,erik, calarco03}
\beq
J(\w)=\alpha \w^3 \e^{-(\w/\w_c)^2},
\label{spectral_density}
\eeq
giving $B=\exp[-(1/2)\int_0^{\infty}d\omega(J(\omega)/\omega^2)\coth{(\beta\omega/2)}].$
The coupling constant $\alpha$ (here having units of $s^2$) captures the strength of the exciton-phonon interaction and 
is dependent upon bulk quantities of the QD sample~\cite{ramsay10}. The exponential cut-off with frequency $\w_c$ arises from the form-factor of the carrier wavefunctions~\cite{krummheuer02, calarco03}. For excitons in 
self-assembled QDs it is proportional to the inverse of the carrier localisation length, which for simplicity we assume to be the same for both electrons and holes.

It is important to note that, apart from the rotating-wave 
approximation on the driving, we have made no further approximations in our manipulations leading 
from Eq.~(\ref{HRWA}) to Eq.~(\ref{eqn:hpolpauli}). We have simply put the Hamiltonian into a form that 
clearly separates the effects of the QD-phonon coupling into renormalisation of QD parameters, through 
$\Omega_r(t)$ and $\omega_0'$, and bath-induced fluctuations, through the last term in $H_{\rm P}$.

\section{Master equation derivation}
\label{master_equation}

Utilising our transformed representation of the QD Hamiltonian, we shall now derive a master equation describing the driven QD exciton dynamics under the influence of the phonon environment. To proceed, we separate the polaron-transformed Hamiltonian such that $H_{\rm P}(t)=H_{\rm 0P}(t)+H_{\rm IP}(t)$. Here, $H_{\rm 0P}(t)=H_{\rm SP}(t)+H_{\rm BP}$, with bath Hamiltonian $H_{\rm BP}=\sum_{\bf k}\omega_{\bf k}b_{\bf k}^{\dagger}b_{\bf k}$, and time-dependent system part
\beq
H_{\rm SP}(t)=\frac{\delta'}{2}\sigma_z+\frac{\Omega_r(t)}{2}\sigma_x,
\eeq
while
\begin{equation}
H_{\rm IP}(t)=\frac{\Omega(t)}{2}\left(\sigma_xB_x+\sigma_yB_y\right)
\end{equation}
is the interaction Hamiltonian, to be treated as a perturbation. Moving into the interaction picture with respect to $H_{\rm 0P}(t)$ yields an interaction Hamiltonian in the (polaron-transformed) interaction picture of the form 
\begin{equation}\label{tdependentHintpol}
\tilde{H}_{\rm IP}(t)=U_{\rm 0P}^{\dagger}(t)H_{\rm IP}(t)U_{\rm 0P}(t),
\end{equation}
where $U_{\rm 0P}(t)=U_{\rm SP}(t)e^{-iH_{\rm BP}t}$, with 
\begin{equation}\label{Uzero}
U_{\rm SP}(t)=\mathrm{T}\exp{\left[-i\int_{0}^tdvH_{\rm SP}(v)\right]}.
\end{equation}
Here, the Schr\"odinger and interaction pictures have been chosen to coincide at time $t=0$, while the time-ordering operator $\mathrm{T}$ is necessary as, in general, $H_{\rm SP}(t)$ does not commute with itself at two different times. We therefore write the interaction Hamiltonian as
\begin{equation}\label{tdependentHintxy}
\tilde{H}_{\rm IP}(t)=\frac{\Omega(t)}{2}\left(\tilde{\sigma}_x(t)\tilde{B}_x(t)+\tilde{\sigma}_y(t)\tilde{B}_y(t)\right),
\end{equation}
where $\tilde{\sigma}_l(t)=U_{\rm SP}^{\dagger}(t)\sigma_lU_{\rm SP}(t)$ and $\tilde{B}_l(t)=e^{iH_{\rm BP}t}B_le^{-iH_{\rm BP}t}$,
for $l=x,y$.

We now follow the standard projection-operator procedure, outlined in Ref.~\cite{b+p}, to derive a time-local master equation for the reduced system density operator, $\tilde{\rho}_{\rm SP}(t)$, in the polaron frame interaction picture. Considering the QD to be initialised in its ground state, with the bath initially in thermal equilibrium, $\rho_{\rm B}(0)=\rho_{\rm B}=e^{-\beta\sum_{\bf k}\omega_{\bf k}b_{\bf k}^{\dagger}b_{\bf k}}/{\rm tr}_B(e^{-\beta\sum_{\bf k}\omega_{\bf k}b_{\bf k}^{\dagger}b_{\bf k}})$, we see that the initial system-bath density operator, $\chi(0)=|0\rangle\langle0|\rho_B$, is unaffected by transformation into the polaron representation, i.e. $\chi_{\rm P}(0)=e^S\chi(0)e^{-S}=|0\rangle\langle0|\rho_B=\chi(0)$. Hence, taking a thermal equilibrium state of the bath as a reference state and treating $\tilde{H}_{\rm IP}(t)$ to second order, we find a homogeneous equation~\cite{b+p, jang08}
\begin{equation}
\label{TCL2inhom}
\frac{\partial\tilde{\rho}_{\rm SP}(t)}{\partial t}=-\int_0^tds{\rm tr}_B\big([\tilde{H}_{\rm IP}(t),[\tilde{H}_{\rm IP}(s),\tilde{\rho}_{\rm SP}(t)\rho_B]]\big),
\end{equation}
describing the dynamics of the excitonic system in the polaron frame, under the influence of the phonon bath.
Substituting in from Eq.~(\ref{tdependentHintxy}), we obtain
\begin{eqnarray}
\label{eqn:intpictmaster}
\frac{\partial\tilde{\rho}_{\rm SP}(t)}{\partial t}=-\frac{\Omega(t)}{4}\int_0^tds\Omega(s)&{}\Big({}&[\tilde{\sigma}_x(t),\tilde{\sigma}_x(s)\tilde{\rho}_{\rm SP}(t)]\Lambda_{x}(\tau)\nonumber\\
&&\:{+}[\tilde{\sigma}_y(t),\tilde{\sigma}_y(s)\tilde{\rho}_{\rm SP}(t)]\Lambda_{y}(\tau)+{\rm H.c.}\Big),
\end{eqnarray}
where ${\rm H.c.}$ refers to the Hermitian conjugate, and we have made use of the stationarity of the bath reference state to write
\beq
\langle \tilde{B}_l(t)\tilde{B}_l(s)\rangle_B=\langle \tilde{B}_l(\tau)\tilde{B}_l(0)\rangle_B=\Lambda_{l}(\tau),
\eeq
with $\tau=t-s$. We note also that owing to the form of $B_x$ and $B_y$, correlation functions of the type 
$\langle \tilde{B}_l(\tau)\tilde{B}_{l'}(0)\rangle_B$ for $l\neq l'$ are identically equal to zero.
Using $J(\w)=\sum_\kv |g_\kv |^2\delta(\w-\w_\kv)$ allows us to write the relevant correlation functions in the continuum limit as
\begin{eqnarray}
\Lambda_{x} (\tau)&{}={}&\frac{B^2}{2}(\e^{\phi(\tau)}+\e^{-\phi(\tau)}-2),\label{Lambdax}\\
\Lambda_{y} (\tau)&{}={}&\frac{B^2}{2}(\e^{\phi(\tau)}-\e^{-\phi(\tau)}),\label{Lambday}
\end{eqnarray}
where
\beq
\phi(\tau)=\int_0^{\infty}d\w\frac{J(\w)}{\w^2}\Bigl(\cos \w\tau\coth(\beta\w/2)-i \sin \w\tau\Bigr).
\label{phi}
\eeq
Now, moving back into the Schr\"{o}dinger picture, and 
making the change of variables $s\rightarrow t-\tau$, we obtain
\begin{eqnarray}
\label{eqn:schrpictmaster2}
\dot{\rho}_{\rm SP}(t)&{}={}&-\frac{i}{2}[\delta' \sigma_z+\Omega_r(t)\sigma_x,\rho_{\rm SP}(t)]\nonumber\\
&&\:{-}\frac{\Omega(t)}{4}\int_0^{t}d\tau\Omega(t-\tau)\bigg([{\sigma}_x,\sigma_x(t-\tau,t){\rho}_{\rm SP}(t)]\Lambda_{x}(\tau)\nonumber\\
&&\:{+}[{\sigma}_y,\sigma_y(t-\tau,t){\rho}_{\rm SP}(t)]\Lambda_{y}(\tau)+{\rm H.c.}\bigg),
\end{eqnarray}
where 
$\sigma_l(s,t)=U_{\rm SP}(t)U_{\rm SP}^{\dagger}(s)\sigma_lU_{\rm SP}(s)U_{\rm SP}^{\dagger}(t)$. 
Eq.~({\ref{eqn:schrpictmaster2}}) is a non-Markovian master equation describing the QD exciton dynamics in the polaron frame for a time-dependent laser-driving pulse envelope $\Omega(t)$, and valid to second order in $H_{\rm IP}(t)$. 

\subsection{Markov approximation}
\label{markov_approximation}

While we could directly use Eq.~(\ref{eqn:schrpictmaster2}) as a basis for numerical simulation of the exciton dynamics (and we shall in fact do so in Section~\ref{nonmarkov}), a great deal of insight into the system behaviour can be gained through the simplifications allowed by the Markov approximation. To make a Markov approximation in the present case, we let the upper limit of integration in Eq.~({\ref{eqn:schrpictmaster2}}) go to infinity under the assumption that the bath correlation functions $\Lambda_{l}(\tau)$ decay on a timescale that is short compared to that of the system dynamics we would like to capture. Given this, we may also approximate
\begin{equation}
U_{\rm SP}(t-\tau,t)\approx\exp{\left[iH_{\rm SP}(t)\tau\right]},
\end{equation} 
while replacing $\Omega(t-\tau)$ by $\Omega(t)$ in the integral in Eq.~(\ref{eqn:schrpictmaster2}). 
We may then write 
\begin{equation}\label{sxt}
\sigma_x(t-\tau,t)\approx\frac{\delta'^2\cos{\eta\tau}+\Omega_r(t)^2}{\eta^2}\sigma_x
+\frac{\delta'\sin{\eta\tau}}{\eta}\sigma_y
+\frac{\delta'\Omega_r(t)(1-\cos{\eta\tau})}{\eta^2}\sigma_z,
\end{equation}
and
\begin{equation}\label{syt}
\sigma_y(t-\tau,t)\approx-\frac{\delta'\sin{\eta\tau}}{\eta}\sigma_x+\cos{\eta\tau}\sigma_y+\frac{\Omega_r(t)\sin{\eta\tau}}{\eta}\sigma_z,
\end{equation}
where $\eta=\sqrt{\delta'^2+\Omega_r(t)^2}$. 

In the following, we shall consider the case of resonant excitation, $\delta'=0$, which simplifies Eqs.~({\ref{sxt}}) and~({\ref{syt}}) to $\sigma_x(t-\tau,t)=\sigma_x$ and 
$\sigma_y(t-\tau,t)\approx \cos(\Omega_r(t) \tau)\sigma_y+\sin(\Omega_r(t)\tau)\sigma_z $, respectively.
We then arrive at a polaron transformed master equation
\begin{eqnarray}
\dot{\rho}_{\rm SP}(t)&{}={}&-\frac{i}{2}[\Omega_r(t)\sigma_x,\rho_{\rm SP}(t)]
-\frac{\Omega(t)^2}{4}\int_0^{\infty}d\tau
\Big([{\sigma}_x,\sigma_x \rho_{\rm SP}(t)]\Lambda_{x}(\tau)\nonumber\\
&&\:{+}\cos(\Omega_r(t)\tau)[{\sigma}_y,\sigma_y\rho_{\rm SP}(t)]\Lambda_{y}(\tau)\nonumber\\
&&\:{+}\sin(\Omega_r(t)\tau)[{\sigma}_y,\sigma_z\rho_{\rm SP}(t)]\Lambda_{y}(\tau)+{\rm H.c.}\Big),
\label{eqn:schrpictmastermarkovresonant}
\end{eqnarray}
which we shall use to explore the dynamics of a resonantly driven QD beyond the weak exciton-phonon coupling 
regime. For constant driving, Eq.~({\ref{eqn:schrpictmastermarkovresonant}}) is of a familiar Born-Markov form in the polaron frame.~\footnote{However, upon transformation back in the original (or ``lab") frame changes in the QD state can 
have an influence on the phonon bath. Specifically, the bath state is not stationary in the original frame, and the system-bath state is not generally separable.}

\subsection{Regimes of validity}
\label{svalidity}

Having derived Eqs.~(\ref{eqn:schrpictmaster2}) and~({\ref{eqn:schrpictmastermarkovresonant}}) perturbatively in the polaron frame, we should expect their validity to be limited in some manner. Recall that the polaron transformation displaces the bath oscillators in reaction to a change of state of the QD. Intuitively, we would therefore expect the polaron transformed representation to be applicable when the bath is able to react on a timescale shorter than, or similar to, that on which the QD exciton itself evolves. Since the timescale 
on which the bath reacts is set approximately by the inverse of the cut-off frequency ($\tau_{\rm B}\sim1/\w_c$), we would therefore expect Eq.~({\ref{eqn:schrpictmaster2}}) to 
work best in the regime $\Omega/\w_c <1$. Additionally, the Markov approximation made in deriving Eq.~({\ref{eqn:schrpictmastermarkovresonant}}) limits its validity to timescales greater than $\tau_{\rm B}$.

To put these considerations on a slightly more quantitative footing, we can make a rough estimate of the regime of validity of our perturbative expansion by considering the magnitude of the perturbative terms in the master equation, namely $(\Omega^2/4)\Lambda_l(\tau)$~\cite{b+p, jang08}, for constant driving $\Omega(t)=\Omega$. For example, consider the upper bound on the magnitude of $\Lambda_y(\tau)$, given by $|\Lambda_y(0)|=(1/2)(1-B^4)$. Bearing in mind that $\Lambda(\tau)$ tends to zero on a timescale of order $1/\omega_c$, we see that we want $(\Omega^2/4)(|\Lambda_y(0)|/\omega_c)=(\Omega^2/8\omega_c)(1-B^4)$ to be small in the sense that terms higher than second order in $H_{\rm IP}$ may be neglected in the master equation expansion. Since $\langle H_{\rm IP}\rangle=0$, the next term is of fourth order, and its magnitude can be estimated in a similar manner by $(\Omega^2/4)^2(|\Lambda_y(0)|^2/\omega_c^3)$. Thus, ignoring numerical factors, we find that the fourth order term is small in comparison to the second order, provided that the condition
\begin{equation}\label{validity}
\left(\frac{\Omega}{\omega_c}\right)^2(1-B^4)\ll 1,
\end{equation}
is satisfied~\footnote{Considering instead the magnitude of $\Lambda_x(\tau)$ leads to a similar condition, which gives essentially the same regime of validity.}. In line with our previous intuition, this condition tells us that in the scaling limit ($\Omega/\omega_c\ll 1$) we expect our treatment to be valid well beyond a standard weak system-bath coupling approach, such that we can explore both the weak ($B\approx 1$, small $\alpha$ and/or low $T$) and strong ($B\ll 1$, large $\alpha$ and/or high $T$) system-bath coupling regimes, as well as reliably interpolate between these two extremes~\cite{wurger98}. Outside the scaling limit, our approach should remain valid provided that the system-bath coupling is small enough, or the temperature low enough, such that the inequality of Eq.~(\ref{validity}) is still satisfied. 

To demonstrate how the polaron approach generally allows a larger regime of parameter space to be explored than a standard weak-coupling treatment, we can also apply the above reasoning to assess the regime of validity of such a weak-coupling master equation. In this case, we have a weak-coupling correlation function $\Lambda_W(\tau)=\int_0^{\infty}d\omega J(\omega)(\cos{(\omega\tau)}\coth{(\beta\omega/2)}-i\sin{(\omega\tau)})$~\cite{nazir08}, which again falls to zero on a timescale of order $1/\omega_c$. Hence, in a similar manner to before, we estimate the second order perturbation to be roughly of magnitude $|\Lambda_W(0)|/\omega_c$, while the fourth order is then $|\Lambda_W(0)|^2/\omega_c^3$. We then find the condition
\begin{equation}\label{validityweak}
\frac{|\Lambda_W(0)|}{\omega_c^2}\ll1,
\end{equation}
as an estimate of the range of validity of the weak-coupling approach.

Considering first the zero temperature limit, we find $\Lambda_W(0)\sim\alpha\omega_c^4$, where we use the QD spectral density given in Eq.~(\ref{spectral_density}). Hence, at zero temperature, our condition implies that a weak exciton-phonon coupling treatment should be adequate to describe the QD excitonic dynamics provided that
\begin{equation}\label{validityweakzerotemp}
\alpha\omega_c^2\ll1.
\end{equation} 
However, as temperature is increased, the magnitude of $\Lambda_W(0)$ does too, and we therefore expect the weak-coupling treatment to worsen. Approximating $\coth{(\beta\omega/2)}\approx2/(\beta\omega)$, we find  
\begin{equation}\label{validityweakhightemp}
\frac{\alpha\omega_c}{\beta}\ll1,
\end{equation} 
or $\alpha\omega_c^2/(\beta\omega_c)\ll1$, which is clearly a harder criterion to fulfill than the zero temperature condition. Hence, for a given system-bath coupling strength $\alpha$ and cutoff frequency $\omega_c$, as the temperature of the bath is increased, a weak-coupling treatment of the system-bath interaction becomes a worse approximation.

Though we should be wary of reading too much into numerical values obtained from these rough validity conditions, for the system studied in Refs.~\cite{ramsay10,ramsay10_2} we can take $\alpha=0.027$~ps$^2$ and $\omega_c=2.2$~ps$^{-1}$ extracted through fits to the data, which gives $\alpha\omega_c^2\approx0.1$. Hence, we might expect a weak-coupling treatment to be valid at low temperatures for this QD system, as borne out by the excellent agreement between experimental data and theory in Refs.~\cite{ramsay10,ramsay10_2}. However, by a temperature of $50$~K, we find $|\Lambda_W(0)|/\omega_c^2\approx0.4$, such that the weak-coupling approximation is now becoming dubious. In contrast, for the same parameters, and taking $\Omega=1$~ps$^{-1}$, we find that the polaron condition [Eq.~(\ref{validity})] gives $(\Omega/\omega_c)^2(1-B^4)\approx0.03$ at $T=0$, increasing up to $(\Omega/\omega_c)^2(1-B^4)\approx0.15$ at $T=50$~K. In fact, we shall show below that it is around temperatures of $30$~K and above that we begin to see significant differences between the weak-coupling and polaron treatments of our driven QD, signifying (in this case) that the system is beginning to move out of the weak-coupling regime, and both driving-renormalisation and multiphonon processes are starting to become important.

\section{Resonant excitation dynamics}
\label{dynamics}

We now proceed to explore the excitonic dynamics of our QD system, focusing in particular on comparing how the polaron and weak-coupling theories capture the interplay between the driving-induced coherent population oscillations and the phonon environment as we vary the temperature. Interestingly, we shall see that as the phonon-induced damping rate naturally depends upon the renormalised Rabi frequency $\Omega_r$ in the polaron theory, but on the original Rabi frequency $\Omega$ in the weak-coupling theory, the weak-coupling approach can actually overestimate the damping rate even in the high-temperature regime where mulitphonon effects are important.

Experimentally, it is generally the excitonic population $\rho_{\rm XX}$ that is measured, for example through 
photocurrent detection~\cite{zrenner02,ramsay10} or microcavity-asissted photon emission~\cite{muller07,flagg09}. Here, we express the solutions to our Markovian master equation through the Bloch vector, defined in the polaron frame as $\vec{\alpha}_{\rm P}=(\alpha_{x{\rm P}},\alpha_{y{\rm P}},\alpha_{z{\rm P}})^T=(\langle\sigma_x \rangle_{\rm P}, \langle\sigma_y \rangle_{\rm P}, \langle\sigma_z \rangle_{\rm P})^T$, where $\langle\sigma_i\rangle_{\rm P}={\rm tr}_{S+B}(\sigma_i\chi_{\rm P}(t))$, for $i=x,y,z$. Since $\sigma_z$ is invariant under the polaron transformation, $e^S\sigma_ze^{-S}=\sigma_z$, we see that in the original (lab) frame $\alpha_z={\rm tr}_{S+B}(\sigma_z\chi(t))={\rm tr}_{S+B}(\sigma_z\chi_{\rm P}(t))=\alpha_{z{\rm P}}$, and the Bloch vector elements along $z$ are equivalent in the two representations. Hence, $\rho_{\rm XX}=(1+\alpha_z)/2=(1+\alpha_{z{\rm P}})/2$, and we may work entirely 
in the polaron frame provided we are only interested in population dynamics. 

From Eq.~(\ref{eqn:schrpictmastermarkovresonant}) we find that the polaron frame Bloch vector evolves according to
\beq\label{labbloch}
\dot{\vec{\alpha}}_{\rm P}=M(t)\cdot\vec{\alpha}_{\rm P}+\vec{b}(t),
\eeq
where
\beq
M(t)=\left(
\begin{array}{ccc}
-(\Gamma_z-\Gamma_y) & 0 & 0 \\
0 & -\Gamma_y & -\Omega_r(t) \\
0 & (\Omega_r(t)+\lambda) & -\Gamma_z
\end{array}
\right),
\eeq
and $\vec{b}(t)=(-\kappa_x,0,0)^T$. Here,
\begin{eqnarray}
\Gamma_y&{}={}&\frac{\Omega(t)^2}{2}\gamma_{x}(0),\label{gammay}\\
\Gamma_z&{}={}&\frac{\Omega(t)^2}{4}\left(\gamma_{y}(\Omega_r(t))+\gamma_{y}(-\Omega_r(t))+2\gamma_{x}(0)\right),\label{gammaz}\\
\lambda&{}={}&\frac{\Omega(t)^2}{2}\left(S_{y}(\Omega_r(t))-S_{y}(-\Omega_r(t))\right),\label{polaron_LS}\\
\kappa&{}={}&\frac{\Omega(t)^2}{4}\left(\gamma_{y}(\Omega_r(t))-\gamma_{y}(-\Omega_r(t))\right)\label{kappa},
\end{eqnarray}
where
\beq
\gamma_{l}(\w)=2 \mathrm{Re}[K_{l}^{\rm (P)}(\w)],\label{gammaresponse}
\eeq
and
\beq
S_{l}(\w)=\mathrm{Im}[K_{l}^{\rm (P)}(\w)],
\eeq
written in terms of the  polaron response function
\beq\label{polresponse}
K_{l}^{\rm (P)}(\w)=\int_0^{\infty}d\tau\e^{i \w \tau}\Lambda_{l}(\tau).
\eeq
Note that as we are solely interested in the exciton population dynamics it suffices to consider only the Bloch equations for $\alpha_y$ and $\alpha_z$ since, in the resonant case, 
that for $\alpha_x$ becomes decoupled.

The polaron theory developed here will be compared to the Born-Markov weak-coupling treatment presented in 
Refs.~\cite{ramsay10, ramsay10_2}. With the weak-coupling theory we again find that the equation of motion for $\alpha_x$ is decoupled, while $\alpha_y$ and $\alpha_z$ obey 
\begin{eqnarray}
\dot{\alpha}_y&{}={}&-\Gamma_W\alpha_y-(\Omega(t)+\lambda_W)\alpha_z,\label{weak_blochy}\\
\dot{\alpha}_z&{}={}&\Omega(t)\alpha_y.
\label{weak_blochz}
\end{eqnarray}
The damping rate and energy shift can be expressed in our current notation as 
\begin{eqnarray}
\Gamma_W&{}={}&\frac{1}{4}\left(\gamma_W(\Omega(t))+\gamma_W(-\Omega(t))\right),\\
\lambda_W&{}={}&\frac{1}{2}\left(S_W(\Omega(t))-S_W(-\Omega(t))\right)\label{weak_lamb_S},
\end{eqnarray}
respectively, with the weak-coupling correlation function given by
\beq
\Lambda_W(\tau)=\int_0^{\infty}J(\w)d\w(\cos\w \tau\coth{(\beta \w/2)}-i\sin\w \tau).
\label{weak_correlation_function}
\eeq
We can evaluate $\Gamma_W$ in closed form, giving 
\beq
\Gamma_W=\frac{\pi}{2}J(\Omega(t))\coth(\beta \Omega(t)/2).
\label{weak_rate}
\eeq
Hence, the weak-coupling rate displays a linear temperature dependence in the high-temperature regime~\cite{ramsay10} and, as mentioned previously, is dependent upon the original Rabi frequency $\Omega(t)$ as opposed to the bath-renormalised value. Furthermore, we see that there is no pure-dephasing contribution to the weak-coupling rate in the Born-Markov approximation, in contrast to the terms $\gamma_x(0)$ appearing in the polaron theory through $\Gamma_y$ and $\Gamma_z$.

We also point out that in the derivation of Eqs.~({\ref{weak_blochy}}) and ({\ref{weak_blochz}}), no secular approximation is made~\cite{b+p}. The corresponding master equation is therefore not of Lindblad form and does not guarantee a completely positive, trace preserving map. As such, for certain parameters, it is possible that Eqs.~({\ref{weak_blochy}}) and ({\ref{weak_blochz}}) predict unphysical behaviour. In fact, we shall see in the following sections that this is indeed the case for high enough temperatures.

\subsection{Time-dependent driving}
\label{time-dependent_driving}

 \begin{figure}[!t]
\begin{center}
\includegraphics[width=0.5\textwidth]{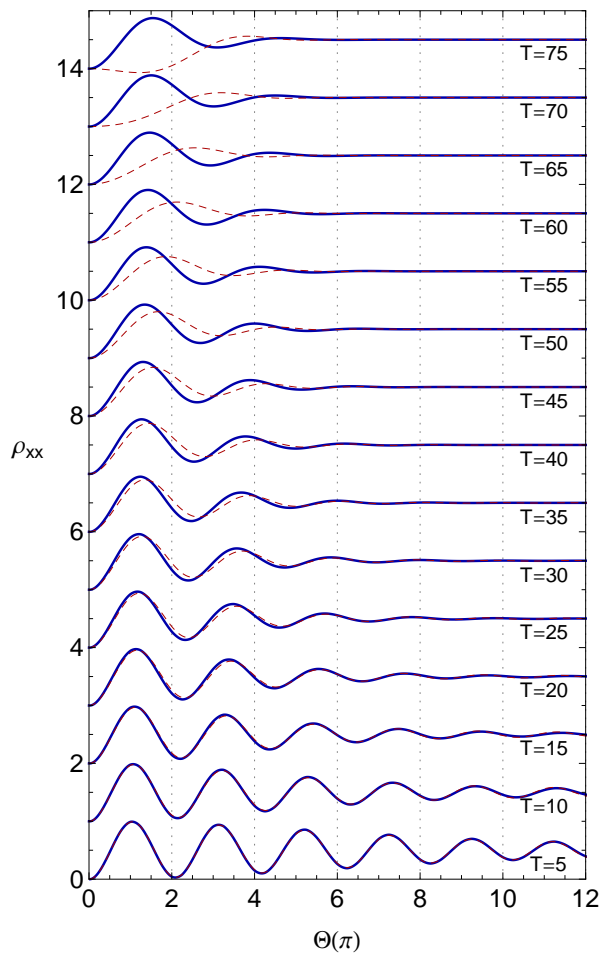}
\caption{Excitonic population as a function of driving pulse area (in units of $\pi$), for temperatures ranging from $5$~K to $75$~K, where each curve has been off-set by an increasing integer for clarity. Blue solid lines are calculated using the polaron approach, while red dashed lines are 
calculated using weak-coupling theory. Parameters: 
$\alpha=0.027~\mathrm{ps}^2$ and $\w_c=2.2~\mathrm{ps}^{-1}$.}
\label{fig:rabi_rotations}
\end{center}
\end{figure}

Having outlined some of the similarities and differences between the weak-coupling and polaron transform approaches, we shall now compare their respective predictions in the case of resonant driving with a Gaussian pulse envelope. Rather than looking at the dynamics in the time domain, we shall instead explore oscillations in the excitonic population (Rabi rotations) as a function of varying {\it pulse area}, $\Theta=\int_{-\infty}^{+\infty} \Omega(t)dt$, for fixed pulse duration, as is common experimentally. We therefore consider a Gaussian pulse of fixed width $\tau$ but varying peak magnitude, centred around $t=0$, and described by $\Omega(t)=(\Theta/2\tau\sqrt{\pi})\mathrm{exp}[-(t/2\tau)^2]$. Starting at a time $-t_0$ well before the pulse (i.e. $t_0\gg \tau$), we initialise the QD in its ground state: $\vec{\alpha}_{\rm P}=\vec{\alpha}=(0,0,-1)^T$. 
We then numerically solve the 
Bloch equations (Eq.~(\ref{labbloch}) in the polaron case, Eqs.~(\ref{weak_blochy}) and~(\ref{weak_blochz}) in the weak-coupling case) to find the state 
of the system at any time $t$ satisfying $t\gg\tau$, such that the pulse has effectively ended. 

Fig.~\ref{fig:rabi_rotations} shows the final excitonic population, $\rho_{\rm XX}$, 
calculated from the polaron and weak-coupling theories as described above, 
as a function of total pulse area, $\Theta$ (in units of $\pi$), for temperatures ranging from $T=5$~K to $T=75$~K (each plot has been offset by an increasing integer for clarity). We use experimentally determined values of the exciton-phonon coupling strength and cut-off frequency, $\alpha=0.027~\mathrm{ps}^2$ and $\w_c=2.2~\mathrm{ps}^{-1}$, respectively~\cite{ramsay10_2}, and a Gaussian driving pulse of width $\tau=4~\mathrm{ps}$. Note that for the largest pulse areas studied here the ratio $\Omega/\w_c$ has a maximum value of $\sim1.2$.

At low to intermediate temperatures ($T<30$~K), we see that the weak-coupling and polaron theories agree very closely in their predictions for the population dynamics, consistent with the excellent agreement found previously between experimental observations and the weak-coupling theory in this regime~\cite{ramsay10, ramsay10_2}. Importantly, the two theories predict almost exactly the same dependence of the Rabi rotation damping rate and frequency shift~\footnote{Note that due to phonon-induced frequency shifts the maxima and minima of the curves in Fig.~\ref{fig:rabi_rotations} are not expected to occur at integer multiples of $\pi$.} on increasing temperature and pulse area, provided the temperature does not increase much above $30$~K. As expected, the phonon-induced damping is strongly driving-dependent, with oscillations becoming almost totally suppressed at high pulse areas for all but the lowest temperatures~\footnote{We do not enter the undamping regime predicted in Ref.~\cite{vagov2007} for the parameters considered here.}.

Perhaps the most striking feature apparent from Fig.~\ref{fig:rabi_rotations}, however, is that the weak-coupling theory tends to {\it overestimate} the damping effect of the phonons at higher temperatures, when compared to the 
polaron theory. In fact, as we shall see below in the case of constant driving, provided $\Omega/\omega_c\sim0.7$ or smaller, the weak-coupling theory predicts a larger damping rate than the polaron theory at all temperatures for the realistic parameters studied here.

At the single-phonon level, this difference 
can be attributed directly to the temperature-dependent suppression of the driving pulse that occurs in the polaron transformed Hamiltonian (see Eq.~(\ref{eqn:hpolpauli})). 
Among other things, this has the consequence that the rates appearing in the polaron Bloch equations are to be evaluated at the (smaller) renormalised pulse strength $\Omega_r(t)$, rather than at the bare pulse strength $\Omega(t)$ as in the weak-coupling theory. The resulting effect can be seen clearly by expanding the relevant polaron rates $\Gamma_y$ and $\Gamma_z$ (in Eqs.~(\ref{gammay}) and~(\ref{gammaz}), respectively) up to their single-phonon terms. We then find a damping rate of precisely the same form as in the weak-coupling theory,
\begin{eqnarray}
\Gamma_{\rm 1-ph}&{}={}&\frac{\pi}{2}J(\Omega_r(t))\coth\left(\beta\Omega_r(t)/2\right),
\label{eqn:polsinglephonon}
\end{eqnarray}
though evaluated at the renormalised $\Omega_r(t)$, as expected. For low pulse areas, we can approximate $\Gamma_{\rm 1-ph}\approx(\alpha\pi/2)\Omega_r(t)^3\coth\left(\beta\Omega_r(t)/2\right)$. Hence, for single-phonon processes at least, the lessening of the damping rate in the polaron theory is simply due to the fact that we are sampling the spectral density at a lower frequency, since $\Omega_r(t)<\Omega(t)$. Any differences would then become more pronounced at higher temperatures, since this is when $\Omega_r$ most differs from $\Omega$. 

In the full polaron theory, however, the situation is of course much more complicated than this simple 
analysis would suggest. To begin with, we have no particular reason to expect the single-phonon rate of Eq.~(\ref{eqn:polsinglephonon}) to be valid over a larger temperature range than the weak-coupling rate of Eq.~(\ref{weak_rate}), so the sampling of the spectral density at different frequencies in the two theories cannot be the whole story. Looking again, for example, at the full polaron rate $\Gamma_y$ (which in fact disappears in the single-phonon approximation), we see from Eqs.~(\ref{Lambdax}),~(\ref{gammay}),~(\ref{gammaresponse}) and~(\ref{polresponse}) that it may be written
\begin{eqnarray}\label{gammayanalysis}
\Gamma_y=\Omega(t)^2{\rm Re}\int_0^{\infty}d\tau\Lambda_x(\tau)=\frac{\Omega_r(t)^2}{2}{\rm Re}\int_0^{\infty}d\tau(\e^{\phi(\tau)}+\e^{-\phi(\tau)}-2).
\end{eqnarray}
Thus, in determining the overall size of the full polaron rates at higher temperatures, there additionally exists a competition between the multiphonon effects accounted for by the exponentiation of the phonon propagator $\phi(\tau)$, which increases the rate in comparison to the single-phonon approximation, and the overall factor proportional to $\Omega_r(t)^2$, which again acts to decrease it with increasing temperature.

A further feature to draw out from the comparison presented in Fig.~\ref{fig:rabi_rotations} is that while the polaron theory predicts physical behaviour at all temperatures considered, for the highest temperature ($T=75$~K) the weak-coupling theory actually predicts unphysical behaviour, since $\rho_{\rm XX}$ becomes negative for pulse areas $\Theta\sim \pi$. This behaviour can be related to an overestimate of the phonon-induced frequency shift in the weak-coupling analysis at high temperatures, and will again be discussed in more detail below for the case of constant driving.

\subsection{Constant driving}
\label{constant_driving}

In order to put the arguments outlined in the previous section on a more formal footing, it is helpful to consider 
the dynamics of the QD system for constant driving, in which case an analytic form can be given for the population difference, $\alpha_z=\rho_{\rm XX}-\rho_{00}$. We construct a second-order differential equation for the time evolution of $\alpha_z$ in both the polaron and weak-coupling theories. From Eq.~(\ref{labbloch}) we find for the polaron theory (using $\alpha_{z{\rm P}}=\alpha_z$)
\beq
\ddot{\alpha}_z+(\Gamma_y+\Gamma_z)\dot{\alpha}_z+(\Omega_r(\Omega_r+\lambda)+\Gamma_y\Gamma_z)\alpha_z=0,
\label{eqn:polaronsigmaz}
\eeq
which has solution (for $\alpha_z(0)=-1$)
\beq
\alpha_z(t)=-\e^{-\Gamma_{\rm P} t/2}\left(\cos(\xi_{\rm P} t/2)+\frac{\Gamma_{\rm P}}{\xi_{\rm P}}\sin(\xi_{\rm P} t/2)\right),
\label{alpha_z_pol}
\eeq
with damping rate
\begin{equation}\label{GammaP}
\Gamma_{\rm P}=\Gamma_y+\Gamma_z=\frac{\Omega^2}{4}(\gamma_{y}(\Omega_r)+\gamma_{y}(-\Omega_r)+4\gamma_{x}(0)),
\end{equation}
and oscillation frequency
\beq
\xi_{\rm P}=\sqrt{4\Omega_r(\Omega_r+\lambda)-(\Gamma_z-\Gamma_y)^2}.
\label{polaron_frequency}
\eeq
On the other hand, Eqs.~(\ref{weak_blochy}) and~(\ref{weak_blochz}) give for the weak-coupling theory
\beq
\ddot{\alpha}_z+\Gamma_W\dot{\alpha}_z+\Omega(\Omega+\lambda_W)\alpha_z=0,
\label{eqn:weaksigmaz}
\eeq
which has a solution of exactly the same form,
\beq
\alpha_{zW}(t)=-\e^{-\Gamma_W t/2}\left(\cos(\xi_W t/2)+\frac{\Gamma_W}{\xi_W}\sin(\xi_W t/2)\right),
\label{alpha_z_weak}
\eeq
though this time with the weak-coupling damping rate $\Gamma_W$ of Eq.~(\ref{weak_rate}), and oscillation frequency
\beq
\xi_W=\sqrt{4\Omega(\Omega+\lambda_W)-\Gamma_W^2}. 
\label{weak_frequency}
\eeq
In the constant driving case, we may therefore directly compare the rate $\Gamma_{\rm P}$ and frequency $\xi_{\rm P}$ in the polaron theory to the weak-coupling expressions $\Gamma_W$ and $\xi_W$, respectively. 

\begin{figure}[!t]
\begin{center}
\includegraphics[width=0.95\textwidth]{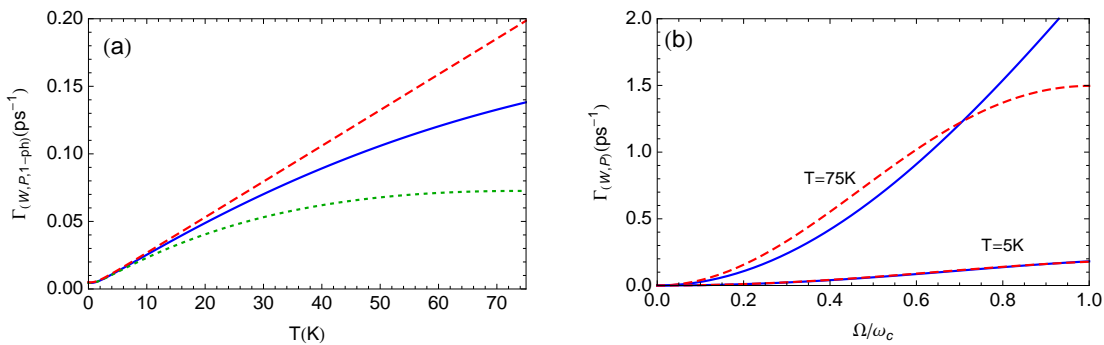}
\caption{(a) Temperature dependence of the weak-coupling rate $\Gamma_W$ (red dashed curve), polaron rate $\Gamma_{\rm P}$ (blue solid curve), and single-phonon expansion of the polaron rate $\Gamma_{\rm 1-ph}$ (green dotted curve), where we have evaluated each 
rate at $\Omega=0.5~\mathrm{ps}^{-1}$. (b) Dependence of the polaron damping rate $\Gamma_{\rm P}$ (blue solid curves) and weak-coupling damping rate $\Gamma_W$ (red dashed curves) on the driving frequency $\Omega$. 
The two sets of curves correspond to temperatures of $T=5$~K and $T=75$~K, as indicated. Parameters: 
$\alpha=0.027~\mathrm{ps}^2$,  $\w_c=2.2~\mathrm{ps}^{-1}$}
\label{fig:rates}
\end{center}
\end{figure}

In Fig.~\ref{fig:rates}(a) we plot the damping rates $\Gamma_{\rm P}$ and $\Gamma_W$, 
along with the single-phonon expansion of the polaron rate ($\Gamma_{\rm 1-ph}$ of Eq.~(\ref{eqn:polsinglephonon})), 
as a function of temperature for an arbitrarily chosen value of the constant driving, $\Omega=0.5~\mathrm{ps}^{-1}$. 
For these parameters, the weak-coupling approximation is indeed shown to predict a larger rate for all values of $T$. 
Furthermore, there is a significant difference between the full polaron rate ($\Gamma_P$) and its single-phonon expansion ($\Gamma_{\rm 1-ph}$) above temperatures of about $10 - 15$~K, indicating that multiphonon effects are 
becoming important. Hence, in this regime, even though the weak-coupling rate is still too large, we cannot simply fix it 
by replacing $\Omega\rightarrow\Omega_r$ in $\Gamma_W$ (i.e. taking $\Gamma_W\rightarrow\Gamma_{\rm 1-ph}$) 
as this neglects important multiphonon processes. Notice also that while the weak-coupling rate varies linearly with temperature above a few Kelvin, the single-phonon expansion does not, despite having a very similar form, due to the temperature dependence inherent to $\Omega_r$.

In Fig.~{\ref{fig:rates}}(b) we show how the polaron and weak-coupling rates vary with the strength of the driving frequency $\Omega$, for low and high temperatures. As observed experimentally~\cite{ramsay10, ramsay10_2}, we see a clear and strong dependence on the driving strength for both temperature regimes, and in both the weak-coupling and polaron theories. It is also interesting to note that around 
$\Omega/\w_c\sim 0.7$ in the high temperature case, the polaron and weak-coupling rates cross, indicating that above this value the hierarchy of rates discussed in reference to Fig.~\ref{fig:rates}(a) no longer holds.

We emphasise again that, as in the case of time-dependent driving, there are two important effects present in the polaron theory which are not captured by the weak-coupling treatment, and which become increasingly relevant as the temperature is increased. 
Firstly, there are multiphonon contributions, which tend to increase the damping rate, as can clearly be seen by comparing the full polaron rate to its single-phonon expansion in Fig.~{\ref{fig:rates}}(a). Secondly, the interaction of the QD exciton with the phonon bath causes a reduction in the effective driving field. For $\Omega/\omega_c<1$, this tends to decrease the damping rate, as can be seen from Fig.~\ref{fig:rates}(b).

\begin{figure}[t!]
\begin{center}
\includegraphics[width=0.99\textwidth]{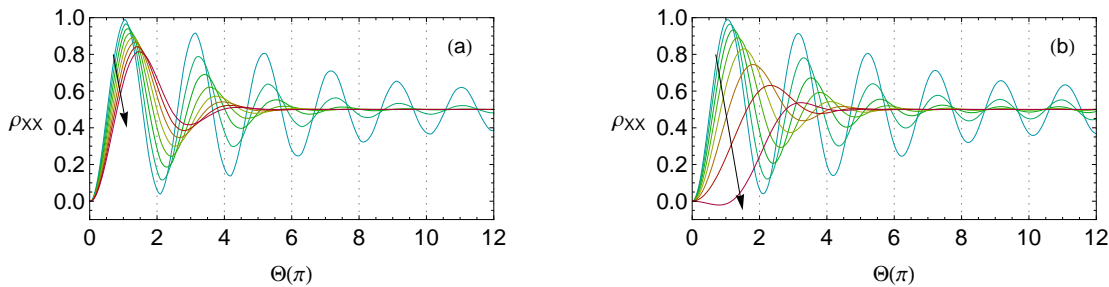}
\caption{Exciton population as a function of pulse area for constant driving: (a) using polaron theory; (b) using weak-coupling theory. The different curves in each plot correspond to temperatures ranging from $5$~K to $75$~K in steps of $10$~K, with lower temperatures coloured blue and higher temperaturea red (the arrows indicate increasing temperature). Parameters: pulse duration $=14~\mathrm{ps}$, $\alpha=0.027~\mathrm{ps}^2$ and $\w_c=2.2~\mathrm{ps}^{-1}$.}
\label{fig:ontopgraphs}
\end{center}
\end{figure}

We are also now in a position to explain the origin of the unphysical behaviour predicted by the weak-coupling theory at $75$~K (see Fig.~{\ref{fig:rabi_rotations}}). In Fig.~{\ref{fig:ontopgraphs}} we again plot the excitonic population as a function of pulse area, but this time for a constant driving pulse of $14$~ps duration, which is roughly equal to the full-width-half-maximum (FWHM) of the Gaussian pulse used in Fig.~{\ref{fig:rabi_rotations}}. Notice that for $T=75$~K, the excited state again takes on unphysical negative values in the weak-coupling theory for pulse areas $\Theta\sim \pi$ (right-hand plot (b)). To see how this comes about, we must consider the weak-coupling oscillation frequency $\xi_W$ of Eq.~(\ref{weak_frequency}). For $\Gamma_W^2>4\Omega(\Omega+\lambda_W)$, we find that $\rho_{\rm XX}$ can take on negative values when $\lambda_W<-\Omega$, i.e. when the correction to the 
driving frequency is larger than the frequency itself, the weak-coupling theory discussed here breaks down. 
As mentioned previously, this can ultimately be attributed to the fact that no secular approximation was made in 
the derivation of the weak-coupling Bloch equations, which therefore have 
a corresponding master equation which is not of Lindblad form~\cite{b+p}.

Let us now consider the frequency shift in slightly more detail. In the weak-coupling theory we find from 
Eqs.~({\ref{weak_lamb_S}}) and ({\ref{weak_correlation_function}}) that
\beq
\lambda_W=\Omega \mathcal{P}\int_0^{\infty}d\omega\frac{J(\w)\coth(\beta \w/2)}{\Omega^2-\w^2},
\label{weak_lamb-shift}
\eeq
where $\mathcal{P}$ indicates that the Cauchy principal value should be taken, and we have made use of the identity 
$\int_0^{\infty}\e^{i \w s}\mathrm{d}s=\pi \delta(\w)+\mathcal{P}(i/\w)$. Since we expect the weak-coupling theory to break down in the high temperature limit, we can evaluate 
$\lambda_W$ analytically by approximating $\coth(x)\approx x^{-1}$ in the integrand of Eq.~({\ref{weak_lamb-shift}}). In 
doing so, we find
\beq
\lambda_W\approx -\frac{\Omega \sqrt{\pi}\alpha\w_c}{\beta}\Big(1-2(\Omega /\w_c)F(\Omega/\w_c)\Big),
\eeq
where $F(x)=\exp[-x^2]\int_0^x\exp[y^2]\mathrm{d}y$ is the Dawson integral. The condition $\lambda_W<-\Omega$, which determines when we expect unphysical behaviour from the weak-coupling theory, then becomes
\beq
\beta<\sqrt{\pi}\alpha\w_c,
\label{unphysical_condition}
\eeq
where we take the limit $\Omega/\w_c\ll1$. For the parameters of Fig.~{\ref{fig:ontopgraphs}}(b),
we then expect to obtain unphysical behaviour when $T>72$~K, in good agreement with the actual dynamics. We note that while Eq.~({\ref{unphysical_condition}}) 
may give a bound on when the limits of the weak-coupling theory are met, its degree of accuracy may become poor well before this condition is satisfied (see Eq.~(\ref{validityweakhightemp}) and discussion there). 

Turning to the low-temperture regime, where the weak-coupling theory remains physical, we have seen previously that by expanding the full polaron damping rate to its single-phonon terms we may recover the weak-coupling damping rate, though evaluated at a renormalised frequency (compare Eqs.~(\ref{weak_rate}) and~(\ref{eqn:polsinglephonon})). In order to complete the picture, we shall now show a similar equivalence between the polaron and weak-coupling frequency shifts at the single-phonon level. Expanding Eq.~({\ref{polaron_LS}}) to first order in $J(\w)$ we find the single-phonon approximation to the polaron frequency shift, $\lambda\rightarrow\lambda_{\mathrm{1-ph}}$, where
\beq
\lambda_{\mathrm{1-ph}}=\Omega_r^3\mathcal{P}\int_0^{\infty}d\omega\frac{J(\w)}{\w^2}\frac{\coth(\beta \w/2)}{(\Omega_r^2-\w^2)}.
\eeq
However, this is not quite the whole story since, in the polaron 
theory, the driving frequency is shifted both by $\lambda$, and also at the Hamiltonian level through $\Omega\rightarrow\Omega_r$. Ultimately, it is the observables, such as the population difference, that are 
the physically meaningful quantities to consider. Inspection of the frequencies $\xi_{\rm P}$ in Eq.~(\ref{polaron_frequency}) and $\xi_W$ in Eq.~(\ref{weak_frequency}) therefore tells us that we should compare $\Omega_r(\Omega_r+\lambda)$ 
in the polaron theory to $\Omega(\Omega+\lambda_W)$ in the weak-couping theory, as we know that $\Gamma_z-\Gamma_y\approx \Gamma_W$ at the single-phonon level. Expanding $\Omega_r$ to first order in $J(\w)$, we find $\Omega_r(\Omega_r+\lambda)\approx\Omega(\Omega+\Delta\Omega)$ where 
\beq
\Delta\Omega=\Omega\int_0^{\infty}d\omega\frac{J(\w)}{\w^2}\coth(\beta \w/2)\left(\frac{\w^2+\Omega_r^2(B^2-1)}{\Omega_r^2-\w^2}\right).
\label{polaron_correction}
\eeq
Expanding the remaining factors of $B$ and occurrences of $\Omega_r$ to first order in $J(\w)$ we find that 
Eq.~({\ref{polaron_correction}}) reduces to Eq.~({\ref{weak_lamb-shift}}) (i.e. $\Delta\Omega\rightarrow\lambda_W$); in 
the weak-coupling limit, the polaron and weak-coupling theories therefore predict the same correction to the driving frequency.

\subsection{Non-Markovian effects}
\label{nonmarkov}

Finally, we shall now relax the Markov approximation made in section~{\ref{markov_approximation}} to investigate non-Markovian effects on the QD exciton dynamics~\cite{machnikowski04, krugel05,vagov2007, alicki04, mogilevtsev08, forstner03, krugel06}, within the polaron frame~\cite{jang08}. Referring to Eq.~({\ref{eqn:schrpictmaster2}}), and considering the case of constant driving for simplicity, we find that our non-Markovian master equation may be written 
\begin{eqnarray}
\label{NM_master_equation}
\dot{\rho}_{\rm SP}(t)&{}={}&-\frac{i}{2}[\delta'\sigma_z+\Omega_r\sigma_x,\rho_{\rm SP}(t)]\nonumber\\
&&-\frac{\Omega^2}{4}\int_0^{t}d\tau\big([{\sigma}_x,\sigma_x(t-\tau,t){\rho}_{\rm SP}(t)]\Lambda_{x}(\tau)\nonumber\\
&&+[{\sigma}_y,\sigma_y(t-\tau,t){\rho}_{\rm SP}(t)]\Lambda_{y}(\tau)+{\rm H.c.}\big).
\end{eqnarray}
Avoiding the Markov approximation thus corresponds to the introduction of time-dependent rates and energy shifts in our master equation~\cite{b+p}. 

\begin{figure}[t!]
\begin{center}
\includegraphics[width=0.95\textwidth]{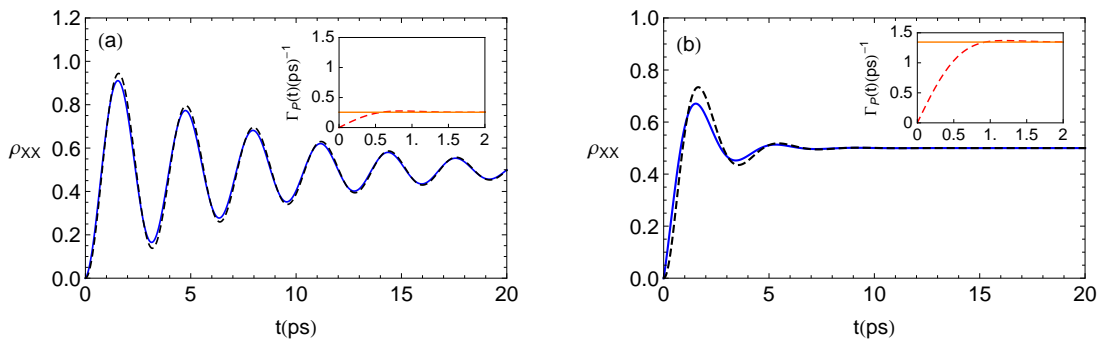}
\caption{Exciton population dynamics in the time domain calculated 
with (solid blue curves) and without (dashed black curves) a Markov approximation. The insets show the non-Markovian decay rates (dashed red curves) approaching their constant 
Markovian values (solid orange lines). Parameters: $\alpha=0.027\mathrm{ps}^2$, $\w_c=2.2\mathrm{ps}^{-1}$, $\Omega=2 \mathrm{ps}^{-1}$. Temperatures: (a) $T=10$~K and (b) $T=50$~K.}
\label{NM_dynamics}
\end{center}
\end{figure}

We determine Bloch equations 
from Eq.~({\ref{NM_master_equation}}) in exactly the same way as the Markovian case. Considering resonant excitation, $\delta'=0$, and inserting $\sigma_x(t-\tau,t)=\sigma_x$ and $\sigma_y(t-\tau,t)= \cos(\Omega_r\tau)\sigma_y+\sin(\Omega_r\tau)\sigma_z$ (which are exact for 
constant driving), we find an equation of motion for the polaron frame Bloch vector 
identical to Eqs.~({\ref{labbloch}})-({\ref{kappa}}) but with $\Omega(t)\rightarrow\Omega$, $\Omega_r(t)\rightarrow\Omega_r$, and all $\gamma_l(\w)$ and $S_l(\w)$ replaced with the time-dependent quantities 
\beq
\gamma_l(\w,t)=2 \mathrm{Re}\bigg[\int_0^t\e^{i \w \tau}\Lambda_l(\tau)\mathrm{d}\tau\bigg],
\label{NM_rate}
\end{equation}
and 
\beq
S_l(\w,t)=\mathrm{Im}\bigg[\int_0^t\e^{i \w \tau}\Lambda_l(\tau)\mathrm{d}\tau\bigg],
\label{NM_lamb}
\eeq
respectively. For the model we consider here, the difference between the non-Markovian and 
Markovian polaron frame dynamics is entirely captured in Eqs.~({\ref{NM_rate}}) and ({\ref{NM_lamb}}). The Markov approximation simply corresponds to pushing the upper integration limits to infinity. We can therefore make the immediate observation that we should expect the Markovian and non-Markovian dynamics to deviate most at short times, since this is when $\gamma_l(\w,t)$ differs significantly from $\gamma_l(\w,\infty)$ (and similarly for $S_l(\w,t)$ and $S_l(\w,\infty)$). These deviations should be most pronounced when $\Lambda_l(0)$ is greatest in magnitude, since this maximises the difference between the Markovian and non-Markovian rates (and energy shifts), and decays on a long timescale (set by $1/\w_c$), as this increases the time over which the non-Markovian rates (and energy shifts) reach their Markovian limits. 

To show that this is indeed the case, in Fig.~{\ref{NM_dynamics}} we plot the excitonic population of our 
QD as a function of time (rather than pulse area). For this figure we take the relatively large value of $\Omega=2~\mathrm{ps}^{-1}$, so that the excitonic system evolves appreciably within the phonon bath correlation time, and in (a) consider a low temperature regime of $T=10$~K. For these parameters non-Markovian effects are most pronounced at short times, as expected, though it is generally fairly difficult to distinguish between the two theories, especially beyond $t\sim 10~\mathrm{ps}$. In the inset we plot the 
non-Markovian generalisation of the polaron theory decay rate (see Eq.~({\ref{GammaP}}))
\beq
\Gamma_P(t)=\frac{\Omega^2}{4}\big(\gamma_{y}(\Omega_r,t)+\gamma_{y}(-\Omega_r,t)+4\gamma_x(0,t)\big),
\eeq
which rapidly approaches its Markovian limit on a timescale $\sim1$~ps.

We can enhance short-time non-Markovian effects by considering higher temperatures, as this increases the difference between the Markov and non-Markov rates and energy shifts on the bath correlation timescale (though it does not change the timescale on which the Markov limit is reached). This is shown in Fig.~{\ref{NM_dynamics}}(b), where we again compare Markovian and non-Markovian dynamics, but 
now at the higher temperature of $T=50$~K. Non-Markovian effects are indeed more pronounced at short times in this case, and, as shown in the inset, the Markov and non-Markov rates do differ more significantly at short times, though the Markov limit is again reached on a similar timescale to that at $10$~K. Once more, beyond $5-10$~ps there is very little to distinguish the Markovian and non-Markovian dynamics.

The inclusion of non-Markovian effects within the polaron frame master equation can therefore affect the population dynamics at short times ($\sim5$~ps and below), but makes very little difference on longer timescales. When plotting excitonic Rabi rotations as a function of pulse area, as in Figs.~{\ref{fig:rabi_rotations}} and {\ref{fig:ontopgraphs}}, it is only the \emph{final} exciton population 
which is measured. For the parameters of Fig.~{\ref{fig:ontopgraphs}}, for example, this corresponds to reading out the excited state population after $14~\mathrm{ps}$. Even for the relatively large Rabi frequencies used in 
Figs.~{\ref{NM_dynamics}}(a) and (b), we see that non-Markovian effects are almost negligible on this timescale. 
Furthermore, at larger temperatures, for which short-time non-Markovian effects seem to be more noticeable, the damping is 
more pronounced, so that the steady state is reached sooner. Hence, since short time behaviour is not captured in the pulse 
area plots of Figs.~{\ref{fig:rabi_rotations}} and {\ref{fig:ontopgraphs}}, neither are non-Markovian effects (remember that in 
Fig.~{\ref{fig:rabi_rotations}} the pulse FWHM is close to $14$~ps). In fact, if we plot the exciton population as a function of 
pulse area using our non-Markovian polaron master equation [Eq.~(\ref{NM_master_equation})] for the same parameters as 
Fig.~{\ref{fig:ontopgraphs}}(a), we find that it is almost indistinguishable from the Markov version on the scale shown in that 
figure. Hence, our theory predicts that (for the parameters considered here) in order for polaron frame non-Markovian 
signatures to be evident in pulse-area plots, FWHM pulse durations on the sub $5$~ps timescale should be used, much shorter than those in the experiments performed in Refs.~\cite{ramsay10, ramsay10_2}.

We comment that Markovian and non-Markovian excitonic dynamics could also be explored within a time non-local master equation formalism~\cite{b+p,wurger98}. 
To do so, one would replace the reduced density operator at time $t$, $\tilde{\rho}_{\txt{SP}}(t)$ in Eq.~({\ref{TCL2inhom}}), with its value instead at time $s$. This generally results in a master equation of convolution-type, which is best solved in Laplace space. 
The dynamics is then determined by the properties of the poles and branch cuts of the Laplace transform of $\alpha_z(t)$, which will have contributions corresponding to both Markovian and non-Markovian processes. Approximating the poles as in Ref.~\cite{wurger98} corresponds to including Markovian effects only, and should therefore be equivalent to our Markovian treatment. It would be interesting to relax this approximation, in particular to include branch cut contributions, and compare the resulting dynamics to the time-local approach used in this section, since it is expected that both approaches should approximate the true dynamics to the same degree of accuracy~\cite{b+p}. We note, however, that a time-dependent driving field is likely to cause complications in the time non-local approach, since it would reduce the applicability of the convolution theorem.

\section{Discussion and summary}
\label{summary}

Inspired by recent experimental observations~\cite{ramsay10, ramsay10_2}, we have investigated the excitonic dynamics 
of a resonantly driven QD under the influence of dephasing due to its interactions with an acoustic phonon environment. We have developed a combined polaron transform, time-local master equation approach to the problem, which accounts for non-perturbative effects such as
multiphonon processes and phonon-induced driving renormalisation.
We have also extended the theory to the non-Markovian regime. 
We have found that for low temperatures ($< 30$~K), the weak-coupling theory presented in Refs.~\cite{ramsay10, ramsay10_2} is in excellent agreement with the polaron master equation dynamics. However, as the temperature is increased, we find that the weak-coupling 
treatment begins to overestimate the damping rate, compared to the polaron theory prediction. In fact, it is interesting to note that in Ref.~\cite{ramsay10_2} 
it was reported that a weak-coupling fit to the data slightly overestimates the damping for temperatures $>40$~K, consistent with our findings. For these temperatures, the non-perturbative aspects of the polaron theory are becoming important. Renormalisation of the Rabi frequency tends to decrease the damping rate, while multiphonon processes act to increase it above the single-phonon level (see Eqs.~({\ref{weak_rate}}), ({\ref{eqn:polsinglephonon}}) and ({\ref{GammaP}}) for the weak-coupling rate, the single-phonon approximation to the polaron rate, and the full polaron rate, respectively).
Deviations from the weak-coupling theory should be even more pronounced at higher temperatures (above the highest temperature of $\sim50$~K explored in Ref.~\cite{ramsay10_2}), though other decoherence mechanisms could also come into play in this regime.

We also considered the important role of the energy-shift terms in the weak-coupling and polaron theories. These terms, analogous to the Lamb-shift in energy levels of atomic physics, are responsible for driving and temperature dependent shifts in the exciton population oscillation frequency, as also reported in Ref.~\cite{ramsay10_2}. While, in general, the energy shifts are necessary for a full description of the dynamics, at high temperatures ($>72$~K for the parameters studied here) we find that in the weak-coupling theory they give rise to unphysical behaviour, and therefore set a bound on the applicability of this approach. On the other hand, the polaron theory suffers no such limitation in this regime.

Finally, we explored the role of non-Markovian effects within the polaron frame, and found that they are predominantly a short-time phenomenon for our experimentally relevant parameters. Hence, they should have little bearing on pulse area plots of Rabi rotations if the pulse duration is long on the bath correlation timescale~\cite{stievater01, zrenner02, htoon02, ramsay10}. Note this implies that, under the same excitation conditions, non-Markovian effects are also negligible in pulse area plots in the weak-coupling theory at low temperatures, since the polaron and weak-coupling approaches agree well in this regime. In order to enhance the visibility of non-Markovian effects, shorter duration pulses or longer bath correlation times are required.

\ack

We would like to thank Andrew Fisher, Sougato Bose, Brendon Lovett, Erik Gauger, and Andrew Ramsay for helpful discussions, comments, and suggestions. This research was supported by the EPSRC.\\


\begin{thebibliography}{10}

\bibitem{bayer00}
M.~Bayer, O.~Stern, P.~Hawrylak, S.~Fafard, and A.~Forchel.
\newblock Hidden symmetries in the energy levels of excitonic `artificial
  atoms'.
\newblock {\em Nature}, 405:923, 2000.

\bibitem{banin99}
U.~Banin, Y.~Cao, D.~Katz, and O.~Millo.
\newblock Identification of atomic-like electronic states in {I}ndium {A}rsenide
  nanocrystal quantum dots.
\newblock {\em Nature}, 400:542, 1999.

\bibitem{bimberg98}
D.~Bimberg, M.~Grundmann, and N.~N. Ledentsov.
\newblock {\em Quantum Dot Heterostructures}.
\newblock Wiley, 1998.

\bibitem{bonadeo98}
N.~H. Bonadeo, J.~Erland, D.~Gammon, D.~Park, D.~S. Katzer, and D.~G. Steel.
\newblock Coherent optical control of the quantum state of a single quantum
  dot.
\newblock {\em Science}, 282:1473, 1998.

\bibitem{michler00}
P.~Michler, A.~Kiraz, C.~Becher, W.~V. Schoenfeld, P.~M. Petroff, L.~Zhang,
  E.~Hu, and A.~Imamoglu.
\newblock A quantum dot single-photon turnstile device.
\newblock {\em Science}, 290:2282, 2000.

\bibitem{stievater01}
T.~H. Stievater, X.~Li, D.~G. Steel, D.~Gammon, D.~S. Katzer, D.~Park,
  C.~Piermarocchi, and L.~J. Sham.
\newblock {Rabi} oscillations of excitons in single quantum dots.
\newblock {\em Phys. Rev. Lett.}, 87:133603, 2001.

\bibitem{kamada01}
H.~Kamada, H.~Gotoh, J.~Temmyo, T.~Takagahara, and H.~Ando.
\newblock Exciton {Rabi} oscillation in a single quantum dot.
\newblock {\em Phys. Rev. Lett.}, 87:246401, 2001.

\bibitem{zrenner02}
A.~Zrenner, E.~Beham, S.~Stufler, F.~Findeis, M.~Bichler, and G.~Abstreiter.
\newblock Coherent properties of a two-level system based on a quantum-dot
  photodiode.
\newblock {\em Nature}, 418:612, 2002.

\bibitem{htoon02}
H.~Htoon, T.~Takagahara, D.~Kulik, O.~Baklenov, A.~L.~Holmes Jr., and C.~K.
  Shih.
\newblock Interplay of {Rabi} oscillations and quantum interference in
  semiconductor quantum dots.
\newblock {\em Phys. Rev. Lett.}, 88:087401, 2002.

\bibitem{borri02}
P.~Borri, W.~Langbein, S.~Schneider, and U.~Woggon.
\newblock {Rabi} oscillations in the excitonic ground-state transition of
  {I}n{G}a{A}s quantum dots.
\newblock {\em Phys. Rev. B.}, 66:081306(R), 2002.

\bibitem{ramsay10}
A.~J. Ramsay, A.~V. Gopal, E.~M. Gauger, A.~Nazir, B.~W. Lovett, A.~M. Fox, and
  M.~S. Skolnick.
\newblock Damping of exciton {Rabi} rotations by acoustic phonons in optically
  excited {I}n{G}a{A}s/{G}a{A}s quantum dots.
\newblock {\em Phys. Rev. Lett.}, 104:017402, 2010.

\bibitem{ramsay10_2}
A.~J. Ramsay, T.~M. Godden, S.~J. Boyle, E.~M. Gauger, A.~Nazir, B.~W. Lovett,
  A.~M. Fox, and M.~S. Skolnick.
\newblock Phonon induced {Rabi} frequency renormalisation of optically driven
  single {I}n{G}a{A}s/{G}a{A}s quantum dots.
\newblock {\em Phys. Rev. Lett.} 105:177402, 2010.

\bibitem{muller07}
A~Muller, E.~B Flagg, P~Bianucci, X.~Y Wang, D.~G Deppe, W~Ma, J~Zhang, G.~J
  Salamo, M~Xiao, and C.~K Shih.
\newblock Resonance fluorescence from a coherently driven semiconductor quantum
  dot in a cavity.
\newblock {\em Phys. Rev. Lett.}, 99:187402, 2007.

\bibitem{flagg09}
E.~B. Flagg, A.~Muller, J.~W. Robertson, S.~Founta, D.~G. Deppe, M.~Xiao,
  W.~Ma, G.~J. Salamo, and C.~K. Shih.
\newblock Resonantly driven coherent oscillations in a solid-state quantum
  emitter.
\newblock {\em Nature Phys.}, 5:203, 2009.

\bibitem{wang05}
Q.~Q. Wang, A.~Muller, P.~Bianucci, E.~Rossi, Q.~K. Xue, T.~Takagahara,
  C.~Piermarocchi, A.~H. MacDonald, and C.~K. Shih.
\newblock Decoherence processes during optical manipulation of excitonic qubits
  in semiconductor quantum dots.
\newblock {\em Phys. Rev. B}, 72:035306, 2005.

\bibitem{li03}
X.~Li, Y.~Wu, D.~Steel, D~Gammon, T.~H Stievater, D.~S Katzer, D~Park,
  C~Piermarocchi, and L.~J Sham.
\newblock An all-optical quantum gate in a semiconductor quantum dot.
\newblock {\em Science}, 301:809, 2003.

\bibitem{santori02}
C.~Santori, D.~Fattal, J.~Vuckovic, G.~S. Solomon, and Y.~Yamamoto.
\newblock Indistinguishable photons from a single-photon device.
\newblock {\em Nature}, 419:594, 2002.

\bibitem{stevenson06}
R.~M. Stevenson, R.~J. Young, P.~Atkinson, K.~Cooper, D.~A. Ritchie, and A.~J.
  Shields.
\newblock A semiconductor source of triggered entangled photon pairs.
\newblock {\em Nature}, 439:179, 2006.

\bibitem{flagg10}
E.~B. Flagg, A.~Muller, S.~V. Polyakov, A.~Ling, A.~Migdall, and G.~S. Solomon.
\newblock Interference of single photons from two separate semiconductor
  quantum dots.
\newblock {\em Phys. Rev. Lett.}, 104:137401, 2010.

\bibitem{patel10}
R.~B. Patel, A.~J. Bennett, I.~Farrer, C.~A. Nicoll, D.~A. Ritchie, and A.~J.
  Shields.
\newblock Two-photon interference of the emission from electrically tunable
  remote quantum dots.
\newblock {\em Nature Photon.}, 4:632, 2010.

\bibitem{kroutvar04}
M.~Kroutvar, Y.~Ducommun, D.~Heiss, M.~Bichler, D.~Schuh, G.~Abstreiter, and
  J.~J. Finley.
\newblock Optically programmable electron spin memory using semiconductor
  quantum dots.
\newblock {\em Nature}, 432:81, 2004.

\bibitem{mikkelsen07}
M.~H. Mikkelsen, J.~Berezovsky, N.~G. Stoltz, L.~A. Coldren, and D.~D.
  Awschalom.
\newblock Optically detected coherent spin dynamics of a single electron in a
  quantum dot.
\newblock {\em Nature Phys.}, 3:770, 2007.

\bibitem{berezovsky08}
J.~Berezovsky, M.~H. Mikkelsen, N.~G. Stoltz, L.~A. Coldren, and D.~D.
  Awschalom.
\newblock Picosecond coherent optical manipulation of a single electron spin in
  a quantum dot.
\newblock {\em Science}, 320:349, 2008.

\bibitem{ramsay08}
A.~J. Ramsay, S.~J. Boyle, R.~S. Kolodka J. B.~B. Oliveira, J.~Skiba-Szymanska,
  H.~Y. Liu, M.~Hopkinson, A.~M. Fox, and M.~S. Skolnick.
\newblock Fast optical preparation, control, and readout of a single quantum
  dot spin.
\newblock {\em Phys. Rev. Lett.}, 100:197401, 2008.

\bibitem{press08}
D.~Press, T.~D. Ladd, B.~Zhang, and Y.~Yamamoto.
\newblock Complete quantum control of a single quantum dot spin using ultrafast
  optical pulses.
\newblock {\em Nature}, 456:218, 2008.

\bibitem{atature06}
M.~Atat{\"u}re, J.~Dreiser, A.~Badolato, A.~H{\"o}gele, K.~Karrai, and
  A.~Imamoglu.
\newblock Quantum-dot spin-state preparation with near-unity fidelity.
\newblock {\em Science}, 312:551, 2006.

\bibitem{gerardot08}
B.~D. Gerardot, D.~Brunner, P.~A. Dalgarno, P.~{\"O}hberg, S.~Seidl, M.~Kroner,
  K.~Karrai, N.~G. Stoltz, P.~M. Petroff, and R.~J. Warburton.
\newblock Optical pumping of a single hole spin in a quantum dot.
\newblock {\em Nature}, 451:441, 2008.

\bibitem{bayer01}
M.~Bayer, P.~Hawrylak, K.~Hinzer, S.~Fafard, M.~Korkusinski, Z.~R. Wasilewski,
  O.~Stern, and A.~Forchel.
\newblock Coupling and entangling of quantum states in quantum dot molecules.
\newblock {\em Science}, 291:451, 2001.

\bibitem{gerardot05}
B.~D. Gerardot, S.~Strauf, M~J.~A. de~Dood, A.~M. Bychkov, A.~Badolato,
  K.~Hennessy, E.~L. Hu, D.~Bouwmeester, and P.~M. Petroff.
\newblock Photon statistics from coupled quantum dots.
\newblock {\em Phys. Rev. Lett}, 95:137403, 2005.

\bibitem{unold05}
T.~Unold, K.~Mueller, C.~Lienau, T.~Elsaesser, and A.~D. Wieck.
\newblock Optical control of excitons in a pair of quantum dots coupled by the
  dipole-dipole interaction.
\newblock {\em Phys. Rev. Lett.}, 94:137404, 2005.

\bibitem{robledo08}
L.~Robledo, J.~Elzerman, G.~Jundt, M.~Atat{\"u}re, A.~H{\"o}gele, S.~F{\"a}lt,
  and A.~Imamoglu.
\newblock Conditional dynamics of interacting quantum dots.
\newblock {\em Science}, 320:772, 2008.

\bibitem{stinaff06}
E.~A Stinaff, M~Scheibner, A.~S Bracker, I.~V Ponomarev, V.~L Korenev, M.~E
  Ware, M.~F Doty, T.~L Reinecke, and D~Gammon.
\newblock Optical signatures of coupled quantum dots.
\newblock {\em Science}, 311:636, 2006.

\bibitem{loss98}
D.~Loss and D.~P. DiVincenzo.
\newblock Quantum computation with quantum dots.
\newblock {\em Phys. Rev. A}, 57:120, 1998.

\bibitem{imamoglu99}
A.~Imamoglu, D.~D. Awschalom, G.~Burkard, D.~P. DiVincenzo, D.~Loss,
  M.~Sherwin, and A.~Small.
\newblock Quantum information processing using quantum dot spins and cavity
  qed.
\newblock {\em Phys. Rev. Lett.}, 83:4204, 1999.

\bibitem{troiani00}
F.~Troiani, U.~Hohenester, and E.~Molinari.
\newblock Exploiting exciton-exciton interactions in semiconductor quantum dots
  for quantum-information processing.
\newblock {\em Phys. Rev. B}, 62:2263(R), 2000; 
F.~Troiani, U.~Hohenester, and E.~Molinari.
\newblock High-finesse optical quantum gates for electron spins in artificial
  molecules.
\newblock {\em Phys. Rev. Lett.}, 90:206802, 2003.

\bibitem{biolatti00}
E.~Biolatti, R.~C. Iotti, P.~Zanardi, and F.~Rossi.
\newblock Quantum information processing with semiconductor macroatoms.
\newblock {\em Phys. Rev. Lett.}, 85:5647, 2000; 
E.~Biolatti, I.~D'Amico, P.~Zanardi, and F.~Rossi.
\newblock Electro-optical properties of semiconductor quantum dots: Application
  to quantum information processing.
\newblock {\em Phys. Rev. B}, 65:075306, 2002.

\bibitem{piermarocchi02}
C.~Piermarocchi, Pochung Chen, L.~J. Sham, and D.~G. Steel.
\newblock Optical {RKKY} interaction between charged semiconductor quantum
  dots.
\newblock {\em Phys. Rev. Lett.}, 89:167402, 2002.

\bibitem{pazy03}
E.~Pazy, T.~Calarco, I.~D'Amico, P.~Zanardi, F.~Rossi, and P.~Zoller.
\newblock Spin-based optical quantum computation via {P}auli blocking in
  semiconductor quantum dots.
\newblock {\em Europhys. Lett.}, 62:175, 2003.

\bibitem{nazir04}
A.~Nazir, B.~W. Lovett, S.~D. Barrett, T.~P. Spiller, and G.~A.~D. Briggs.
\newblock Selective spin coupling through a single exciton.
\newblock {\em Phys. Rev. Lett.}, 93:150502, 2004; 
A.~Nazir, B.~W. Lovett, and G.~A.~D. Briggs.
\newblock Creating excitonic entanglement in quantum dots through the optical
  stark effect.
\newblock {\em Phys. Rev. A}, 70:052301, 2004.

\bibitem{economou08}
S.~E. Economou and T.~L. Reinecke.
\newblock Optically induced spin gates in coupled quantum dots using the
  electron-hole exchange interaction.
\newblock {\em Phys. Rev. B}, 78:115306, 2008.

\bibitem{vasanelli02}
A.~Vasanelli, R.~Ferreira, and G.~Bastard.
\newblock Continuous absorption background and decoherence in quantum dots.
\newblock {\em Phys. Rev. Lett.}, 89:216804, 2002.

\bibitem{villas05}
J.~M. Villas-Boas, S.~E. Ulloa, and A.~O. Govorov.
\newblock Decoherence of {Rabi} oscillations in a single quantum dot.
\newblock {\em Phys. Rev. Lett.}, 94:057404, 2005.

\bibitem{machnikowski04}
P.~Machnikowski and L.~Jacak.
\newblock Resonant nature of phonon-induced damping of {R}abi oscillations in
  quantum dots.
\newblock {\em Phys. Rev. B.}, 69:193302, 2004.

\bibitem{krugel05}
A.~Krugel, V.~M. Axt, P.~Machnikowski, and A.~Vagov.
\newblock The role of acoustic phonons for {R}abi oscillation in semiconductor
  quantum dots.
\newblock {\em Appl. Phys. B.}, 81:897, 2005.

\bibitem{vagov2007}
A.~Vagov, M.~D. Croitoru, V.~M. Axt, T.~Kuhn, and F.~M. Peeters.
\newblock Nonmonotonic field dependence of damping and reappearance of {Rabi}
  oscillations in quantum dots.
\newblock {\em Phys. Rev. Lett.}, 98:227403, 2007.

\bibitem{nazir08}
A.~Nazir.
\newblock Photon statistics from a resonantly driven quantum dot.
\newblock {\em Phys. Rev. B}, 78:153309, 2008.

\bibitem{mahan}
G.~D. Mahan.
\newblock {\em Many-Particle Physics}.
\newblock Plenum, 1990.

\bibitem{jacak03}
L.~Jacak, P.~Machnikowski, J.~Krasnyj, and P.~Zoller.
\newblock Coherent and incoherent phonon processes in artificial atoms.
\newblock {\em Eur. Phys. J. D.}, 22:319, 2003.

\bibitem{krummheuer02}
B.~Krummheuer, V.~M. Axt, and T.~Kuhn.
\newblock Theory of pure dephasing and the resulting absorption line shape in
  semiconductor quantum dots.
\newblock {\em Phys. Rev. B.}, 65:195313, 2002.

\bibitem{vagov02}
A.~Vagov, V.~M. Axt, and T.~Kuhn.
\newblock Electron-phonon dynamics in optically excited quantum dots: Exact
  solution for multiple ultrashrot laser pulses.
\newblock {\em Phys. Rev. B.}, 66:165312, 2002.

\bibitem{grodecka07}
A.~Grodecka, C.~Weber, P.~Machnikowski, and A.~Knorr.
\newblock Interplay and optimization of decoherence mechanisms in the optical
  control of spin quantum bits implemented on a semiconductor quantum dot.
\newblock {\em Phys. Rev. B.}, 76:205305, 2007.

\bibitem{erik}
E.~M. Gauger, A.~Nazir, S.~C. Benjamin, T.~M. Stace, and Brendon~W Lovett.
\newblock Robust adiabatic approach to optical spin entangling in coupled
  quantum dots.
\newblock {\em New J. Phys.}, 10:073016, 2008.

\bibitem{alicki04}
R.~Alicki, M.~Horodecki, P.~Horodecki, R.~Horodecki, L.~Jacak, and
  P.~Machnikowski.
\newblock Optimal strategy for a single-qubit gate and the trade-off between
  opposite types of decoherence.
\newblock {\em Phys. Rev. A.}, 70:010501(R), 2004.

\bibitem{mogilevtsev08}
D~Mogilevtsev, A.~P Nisovtsev, S~Kilin, S.~B Cavalcanti, H.~S Brandi, and L.~E
  Oliveira.
\newblock Driving-dependent damping of {R}abi oscillations in two-level
  semiconductor systems.
\newblock {\em Phys. Rev. Lett.}, 100:017401, 2008.

\bibitem{forstner03}
J.~F{\"o}rstner, C.~Weber, J.~Danckwerts, and A.~Knorr.
\newblock Phonon-assisted damping of {R}abi oscillations in semiconductor quantum
  dots.
\newblock {\em Phys. Rev. Lett.}, 91:127401, 2003.

\bibitem{krugel06}
A.~Krugel, V.~M. Axt, and T.~Kuhn.
\newblock Back action of nonequilibrium phonons on the optically induced
  dynamics in semiconductor quantum dots.
\newblock {\em Phys. Rev. B.}, 73:035302, 2006.

\bibitem{b+p}
H.-P. Breuer and F.~Petruccione.
\newblock {\em The Theory of Open Quantum Systems}.
\newblock Oxford University Press, 2002.

\bibitem{wurger98}
A.~Wurger.
\newblock Strong-coupling theory for the spin-phonon model.
\newblock {\em Phys. Rev. B.}, 57:347, 1998.

\bibitem{wilson02}
I.~Wilson-Rae and A.~Imamoglu.
\newblock Quantum dot cavity-QED in the presence of strong electron-phonon interactions
\newblock {\em Phys. Rev. B}, 65:235311, 2002

\bibitem{jang08}
S.~Jang, Y-C. Cheng, D.~R. Reichman, and J.~D. Eaves.
\newblock Theory of coherent resonance energy transfer.
\newblock {\em J. Chem. Phys.}, 129:101104, 2008.

\bibitem{leggett87}
A.~J. Leggett, S.~Chakravarty, A.~T. Dorsey, M.~Fisher, A.~Garg, and
  W.~Zwerger.
\newblock Dynamics of the dissipative two-state system.
\newblock {\em Rev. Mod. Phys.}, 59:1, 1987.

\bibitem{calarco03}
T.~Calarco, A.~Datta, P.~Fedichev, E.~Pazy, and P.~Zoller.
\newblock Spin-based all-optical quantum computation with quantum dots:
  Understanding and suppressing decoherence.
\newblock {\em Phys. Rev. A}, 68:012310, 2003.

\end{thebibliography}
\end{document}